\documentclass[a4paper,12pt,reqno,superscriptaddress,nofootinbib]{revtex4}
\usepackage[centertags]{amsmath}
\usepackage{amsfonts}
\usepackage{amssymb}
\usepackage{amsthm}
\usepackage{newlfont}
\usepackage{stmaryrd}
\usepackage{mathrsfs}
\usepackage{mathtools}
\usepackage{euscript}
\usepackage{graphicx}
\usepackage{enumerate}
\usepackage{todonotes}
\usepackage[normalem]{ulem} 

\usepackage{color}
\usepackage{floatrow}
\usepackage{caption}

\usepackage{tikz}
\usepackage{pgf}
\usetikzlibrary{positioning,fit,calc}
\usetikzlibrary{arrows,automata}
\usepackage{wrapfig}
\usepackage{subfigure}
\usepackage{amscd}
\usepackage{hyperref}

\theoremstyle{plain}

\theoremstyle{definition}

\theoremstyle{remark}

\newcommand{\be}{\begin{equation}}
\newcommand{\en}{\end{equation}}

\newcommand{\bbZ}{{\mathbb Z}}

\newcommand{\opunit}{\text{1}\kern-0.22em\text{l}}

\newcommand{\id}{\textrm{d}}

\DeclareMathAlphabet{\mathpzc}{OT1}{pzc}{m}{it}

\let\oldsqrt\sqrt
\def\sqrt{\mathpalette\DHLhksqrt}
\def\DHLhksqrt#1#2{%
	\setbox0=\hbox{$#1\oldsqrt{#2\,}$}\dimen0=\ht0
	\advance\dimen0-0.2\ht0
	\setbox2=\hbox{\vrule height\ht0 depth -\dimen0}%
	{\box0\lower0.4pt\box2}}

\DeclareMathAlphabet{\mathpzc}{OT1}{pzc}{m}{it}

\def\bea{\begin{eqnarray}}
\def\eea{\end{eqnarray}}
\def\ba{\begin{array}}
	\def\ea{\end{array}}

\begin{document}

\title{Producing suprathermal tails\\ in the stationary velocity distribution}

\begin{abstract}
We revisit effective scenarios for the origin of heavy tails in stationary velocity distributions.  A first analysis combines localization with diffusive acceleration. That gets realized in space plasmas to find the so called kappa-distributions having power law decay at high speeds. There, localization at high energy already takes place for the reversible dynamics, but becomes effective by an active diffusion in velocity space.  A model for vibrating granular gases and giving rise to stretched exponential tails is also briefly discussed, where negative friction is the energizer.  In all cases, the resulting nonMaxwellian velocity distributions are frenetically caused by the dependence on the speed of kinetic parameters.
\end{abstract}

\author{Thibaut Demaerel, Wojciech De Roeck and Christian Maes \\ {\it
Instituut voor Theoretische Fysica, KU Leuven}}
\maketitle

{\it Dedicated to the memory of Christian Van den Broeck.}

\section{Introduction}
How do Maxwellians originate?  Consider a classical gas of $N$ point particles of mass $m$.  We fix the total energy $E$ which we assume is totally kinetic. Every particle has a velocity $\vec{v}_i$ with only constraint that $\sum_i mv_i^2/2= E$.   We ask what fraction of particles has a velocity $\vec{v}_i \simeq \vec{u} \pm \vec{\delta}$ approximately equal to a given $\vec{u}$; the $\delta>0$ gives some small margin.  Denote that proportion by
\[
p(\vec u) := \frac 1{N}\,\big|\{i=1,\ldots,N:\vec{v}_i\in \Delta(\vec{u})\}\big|
\]
where $\Delta(\vec{u})$ is a cube of volume $\delta^3$ around
$\vec{u}\in (\delta \bbZ)^3$. The main observation is that for very large $N$,
\[
p(\vec u) \simeq
\frac{e^{-\beta m |\vec{u}|^2/2}}{(2\pi m k_B T)^{3/2}}
\]
Here $T$ is the absolute temperature (and $\beta = 1/(k_B T)$) for which $E = 3N k_B
T/2$.  More precisely, no matter how small we choose  $\varepsilon>0$, we can make $N$ large enough so that for almost all $(\vec{v}_1,\ldots,\vec{v}_N)$ on the $E$-surface,
\begin{equation}\label{tt}
\Big|p(\vec u) -
\frac{e^{-\beta m |\vec{u}|^2/2}}{(2\pi m k_B T)^{3/2}}\Big|\leq
\varepsilon
\end{equation}
We conclude therefore that in thermal equilibrium the Maxwell distribution is the {\it typical} velocity distribution, as taught by Boltzmann (1896):
\begin{quote}
The Maxwell distribution
is in no way a special singular distribution which is to be contrasted to infinitely many
more non-Maxwellian distributions; rather it is characterized by the fact that by far the
largest number of possible velocity distributions have the characteristic properties of the
Maxwell distribution.
\end{quote}
There are of course also dynamical ``derivations'' of the Maxwell distribution.  We could e.g. use the Boltzmann equation for dilute gases with hard core interaction to find that the Maxwellian is stationary.  Or, for open systems (Brownian motion) we can consider the Langevin equation and derive the Maxwellian as the stationary distribution.  Those derivations are in no way independent but in fact consequences, be it indirectly, of the counting argument above.\\

The above has a relativistic version (in the so called J\"uttner distribution) but that is of no concern for the present paper.  The general question here is to get a simple understanding of the presence of nonMaxwellians in nature.   Going for even more sensation, we wish to understand the possible origin of heavily occupied tails in the velocity distribution of dilute gases or other classical macroscopic systems.  While one may imagine various possible mechanisms we concentrate on a single idea formalized in Section \ref{dac}: the original reversible dynamics shows localization in energy and when combined with active diffusion in velocity space, the high velocity tails get overpopulated.  What matters most is the dependence on the speed of escape rates and of activity parameters, which are addressed as frenetic contributions or non-dissipative aspects in nonequilibrium statistical mechanics, \cite{spring}.\\
In Section \ref{plas} we illustrate the scenario for the origin of power-law decay in space plasmas. Interestingly, the active velocity diffusion there can be related to Taylor dispersion in velocity space \cite{tdif}, and we borrow the insight of Christian Van den Broeck in \cite{tay} to set up a simple model relating it to the telegraph process.   Similarly, in Section \ref{gran} we connect with driven granular gases for the understanding of stretched exponential decay. There we add negative friction to model the macroscopic impact of a vibrating plate which is another version of stochastic acceleration.  Both applications are not original and for each of those cases (multiple) analyses and reasons have been given in the literature.  E.g. in \cite{mas,kap} one finds a convincing explanation of power law tails in the velocity distribution in space plasmas by means of whistler-mode wave stochastic acceleration. For granular gases a very similar analysis is contained in \cite{dbe,pra}.  There may even be other physical explanations, differing in various details. The simplicity of our reasoning may allow to unify those many versions.
The point of departure in the next section are Markov diffusion processes in velocity space.

\section{Fokker-Planck equation}\label{gen}
In view of the counting argument presented in the beginning of the Introduction it is natural to ask how then it can still be the case that suprathermal tails physically arise in velocity distributions.  It must of course be a nonequilibrium effect, where a driving and the compensation in dissipation gives rise to a stationary condition for an open system dynamics that does not satisfy detailed balance.\\
In the simplest of worlds we assume a three-dimensional effective Markov diffusion dynamics in an isotropic and homogeneous environment for independent particles of mass $m$: in Stratonovich convention,
\begin{equation}
\label{lan}
m\dot{\vec v} = -\gamma(v)\,\vec v + \sqrt{2D(v)}\,\vec{\xi_t},\qquad |\vec v|= v\geq 0
\end{equation}
where the functions $\gamma (v)$ (friction or rate at which the particle decelerates) and $D(v)$ (diffusion in velocity space) will need to be derived in some limiting regime or under some approximations as e.g. discussed in the next sections.  The noise $\vec{\xi}_t$ is standard white noise in three dimensions. We assume that the initial velocity distribution is isotropic so that the corresponding Fokker-Planck equation for the density $\rho(\vec v,t), t\geq 0$, with respect to $\id^3\vec v$ is given by 
\begin{equation}\label{fk}
\frac{\partial \rho}{\partial t}(\vec v,t) =
\frac 1{v^2}\frac{\partial}{\partial v}\left(\frac{\gamma(v)}{m}\,v^3\,\rho(\vec v,t)\right) + \frac 1{v^2}\frac{\partial}{\partial v}\left(v^2\,\frac{D(v)}{m^2}\frac{\partial \rho}{\partial v}(\vec v,t)\right)
\end{equation}
The Maxwellian for particles of mass $m$ in a thermal environment at temperature $T$ is the probability density $\rho_\text{eq}(\vec v) = \frac 1{Z}\exp [-mv^2/2k_BT]$.  
It solves the stationary Fokker-Planck equation \eqref{fk}$=0$,
whenever $D(v) = k_B\,T\,\gamma(v) $ (Einstein relation in velocity space).  Note that the dependence on $v$ of the kinetics hidden  in $\gamma(v) = D(v)/(k_BT)$ does not matter at all then for the stationary density.\\
That Einstein relation gets violated, \cite{chr,chrstef}, and the dependence of the parameters $\gamma(v),D(v)$ on $v$ becomes in fact crucial when the environment is out-of-equilibrium.   In the minimal formal version we still consider in this section it means simply that the function $\gamma(v)$ is no longer proportional to $D(v)$, and then, as is easy to verify, we get the following velocity distribution as stationary solution to \eqref{fk}, 
\begin{equation}\label{nonm}
\rho_s(\vec v) \sim \exp -M(v) 
\end{equation}
where $M'(v) = \frac{m\,\gamma(v)\,v}{D(v)}$ or
\begin{equation}\label{mm}
M(v) = m\,\int_0^v\id u\,\frac{\gamma(u)\, u}{D(u)}
\end{equation}
if the integral makes sense. That function $M(v)$ will contain the coupling strength of the particle with its environment and it will depend on further nonequilibrium features of that environment as well.  Note however that the spatial extension of the system is not involved, nor do we employ the position of the particle to apply the nonequilibrium driving; there will be no systematic force but in the shaking of the velocities --- see below.\\
The conclusion on the shape of the speed distribution \eqref{nonm} is directly reached by inspecting the exponent $\alpha$ in the large speed behavior of the integrand in \eqref{mm},
\begin{equation}\label{fatt}
\frac{\gamma(v)}{D(v)} = \frac 1{k_BT}\,\left(\frac{v}{v_0}\right)^{\alpha -1}
\end{equation}
(where $k_BT$ sets the thermal energy scale and $v_0$ is a corresponding reference speed).
If $\alpha = 1$ we get $M(v) \propto v^2/2$ and a Gaussian (thermal) decay in the speed distribution.  Suprathermal tails are found when $\alpha < 1$ (understanding \eqref{fatt} for $v/v_0 \uparrow \infty$).  For $\alpha=0$ we get an exponential decay. The case  $1>\alpha>-1$ implies compressed to stretched exponential decay ending in $\alpha=-1$ where a power-law decay arises in the speed distribution.  For $\alpha < -1$ the set-up does not make sense for obtaining stationary distributions.\\

Mathematically there are two equivalent ways to get away from $\alpha=1$ in \eqref{fatt}, either to modify $D(v)$ or to modify $\gamma(v)$.  Physically, they are arising from different mechanisms possibly, as we discuss in the next sections.

\section{Diffusive acceleration}\label{dac}

The scenario here is to move from the equilibrium dynamics by changing $D(v)$ without changing $\gamma(v)$.  In other words we consider \eqref{fk} in the version of a sum
\begin{equation}\label{fknon}
\frac{\partial \rho}{\partial t}(\vec v,t) =
\frac 1{v^2}\frac{\partial}{\partial v}\left(\frac{\gamma(v)}{m}\,v^2\,\left[v\,\rho(\vec v,t) + \frac{k_BT}{m}\frac{\partial \rho}{\partial v}(\vec v,t)\right]\right) +   \frac 1{v^2}\frac{\partial}{\partial v}\left(v^2\,\frac{B(v)}{m^2}\frac{\partial \rho}{\partial v}(\vec v,t)\right) 
\end{equation}
which is exactly the same equation as \eqref{fk} but with $D(v) = k_BT\gamma(v) + B(v)$ and where the left-hand side of \eqref{fatt} is changed into
\begin{equation}\label{fatleft}
\frac{\gamma(v)}{k_BT \gamma(v) + B(v)} = \frac{1}{k_BT  + B(v)/\gamma(v)}  
\end{equation}
The exponent $\alpha$ in \eqref{fatt} is decided by
\begin{equation}\label{eta}
\frac{B(v)}{\gamma(v)} \propto v^{2\eta}, \quad v\uparrow \infty,\qquad \alpha = 1-2\eta,\quad 0\leq \eta\leq 1 
\end{equation}
to give the stationary nonequilibrium speed distribution \eqref{nonm},
\begin{equation}\label{resu}
\rho_s(\vec v)\,\id^3\vec v = 4\pi \,v^2\,\rho_s(v)\,\id v \propto \frac{\id v}{v^{2\kappa}}\,\exp -b\,v^{2(1-\eta)},\qquad v\uparrow \infty
 \end{equation}
with power-law exponent $\kappa$ getting important (only) for $\eta=1$, and with $b>0$ also being determined by further parameters such as mass, reference temperature and the nature of the nonequilibrium environment.\\

The physics thus enters in two ways, (1) how the friction parameter $\gamma(v)$ behaves for the detailed balance (undriven dynamics) and (2) what sort of $B(v)$ arises from the driving.  Their effects add in the above set-up \eqref{fknon} where the evolution equation has the test particle exposed either to an equilibrium environment at temperature $T$ or to (effectively infinite temperature) diffusive driving. 
We now discuss $\gamma(v)$ and $B(v)$.

We first determine the friction parameter $\gamma(v)$ for large $v$.  A high-energetic particle experiences a friction force $\gamma(v)$ due to scattering events which cause it to lose energy.   To estimate $\gamma(v)$ for large $v$ we therefore need the large energy scattering cross-section  $\sigma(v)$, with relation
\[
\gamma(v) = v \,\sigma(v)
\]
We ignore the angular dependencies; see e.g.~\cite{dbe} for a detailed analysis.  For example, if we have elastic potential scattering for a central potential $V(r) \propto r^{-a}$ with exponent $a>0$, then the cross-section scales as $\sigma(v) \propto v^{-4/a}$, and
\begin{equation}\label{ela}
\gamma(v) \propto v^{1- 4/a}
\end{equation}
 Inelastic scattering may create other dependencies (see also in Section \ref{gran}). To get a positive $\eta$  in \eqref{eta}--\eqref{resu} when $B(v)$ is not growing with $v$, we need $\gamma(v)$ to decay with $v$.  And $\eta$ gets closer to one when $\gamma(v)$ is decaying faster with large $v$.  That means that for high energy collisions the exchange of energy must be small.   Such an effect is often observed and, for obvious reasons, we refer to it as localization, ensuring (relative) stability of high energy 
 regions.  Recent examples and demonstrations are presented in \cite{drh, cuneo, flach, iubini}.   \\ 

Let us now move to $B(v)$.  One must first realize that the addition of the last term in \eqref{fknon} is (only and again) a diffusive approximation; see e.g. \cite{consi,dol,aba}. In particular one should not be forced to think that the true origin of the nonequilibrium condition is literally the connection (through the test particle) of the finite temperature dynamics (first two terms in \eqref{fknon}) with an infinite temperature reservoir (last term in \eqref{fknon}). On a more mesoscopic scale we must try to imagine the effect on a single particle of time- and space-dependent force fields arising in a turbulent plasma. To estimate $B(v)$ let us disconnect the last term in \eqref{fknon} from the other terms, and only consider the diffusion with amplitude $B$. We need to know how spacetime force fields in the system accelerate the particle (and would have it run away if no sufficient friction is present).  We will treat an example in the next section.  Here it suffices to say that as always
for Markov diffusions, here the last term in \eqref{fknon}, we find the $B(v)$ from the small time ($=\epsilon$) variance
	\begin{equation}\label{md}
m^2\,\langle \left| \vec v(\epsilon) - \vec v(0)\right|^2\,| \vec v(0) = \vec v\rangle =	B(v)\,\epsilon,\qquad \epsilon \downarrow 0
	 \end{equation}
In the left-hand side we condition on the initial velocity to be $\vec v$.
	 If we take $B(v) \simeq v^{-k}$ for some $k$, then for large times $t$,
	 \[
	 |\vec v(t)| \simeq t^{\frac{1}{2+k}}, \quad \text{ or }\;\frac{\id}{\id t}\langle v\rangle  \propto\,\langle v^{-1-k}\rangle 
	 	 \]
Diffusion in velocity space is indeed a form of (stochastic) acceleration, where
 $B(v)$ gives the growth of kinetic energy in time.	 The idea therefore is that the extra diffusion term in \eqref{fknon} injects energy at rate $B(v)$.
 Shaking, vibrating or the presence of time-dependent interactions and fields can all contribute to that when treated in the underdamped case. There is in fact a large literature on the subject which originates from fundamental questions in the foundations of statistical mechanics; see e.g. \cite{pus} and the references to the Ulam model \cite{ula} and to Fermi acceleration \cite{Fer}. Specific analyses have been carried out in a large number of cases; see e.g. \cite{dol,stur,deb1,deb2}. 

The conclusion is that, combined with $\gamma(v)$, the mechanism of energy injection and how it depends on large $v$ determines the exponent $\eta$ in \eqref{eta}--\eqref{resu}.

\section{Power-law tails in space plasmas}\label{plas}
\
The formal set-up of Section \ref{dac} applies to space plasmas.  There one observes suprathermal tails with $\eta=0$, i.e., power law decay in the speed distribution.  That class of heavy tailed distributions are called $\kappa-$distributions.  Discussions on the $\kappa-$distribution in space plasmas have a long history, see e.g. \cite{mas,kap,has} for theoretical discussions.  A review of the various aspects in the detection, origin and consequences of $\kappa-$distributions is found in \cite{peir}.\\

Confronting it with the Section \ref{dac} it is clear that because of the Coulomb scattering $a=1$ in \eqref{ela} (Rutherford scattering). On the other hand, various arguments have been developed arguing that due to some form of electromagnetic driving $B(v) \propto 1/v$; see e.g.~\cite{bak,stix}.   Then, in \eqref{mm}, for large speeds $M(v)= 2(1 + \kappa)\log (\frac{v}{\sqrt{\kappa}\,w})$  in terms of a parameter $\kappa>0$.  That leads to the $\kappa$-(non-thermal)-(nonMaxwellian)-(Lorentzian) velocity distribution which is indeed observed in space plasmas.\\

The plasma physics underlying the above scaling relations is complicated (and even today still somewhat controversial; see e.g. \cite{bern})  especially when dealing with anisotropies and the different types of plasma waves.  Yet, that $B(v) \propto 1/v$ should not be too surprising for a force field when the effective force on the particle averages out to zero in the $v\uparrow \infty$ limit; the first correction around zero as governed by the central limit theorem naturally gives the $1/v$-behavior, see also below for more details. A quantitative analysis can be found in \cite{deb1}, when the kinetic energy grows like $t^{2/3}$ in time.  We propose to see that as a generalization of Fermi-acceleration, which was first discussed in the context of cosmic rays \cite{Fer,dol,pus}, but is known also more broadly as turbulent or diffusive acceleration \cite{deb1,deb2,dol,pus}.  Indeed, Fermi's motivation was to understand the abundance of high energy particles in the cosmos, which is very related of course to the subject of suprathermal tails. \\

Let us now sketch two concrete models from which the scaling $B(v)\sim 1/v$ emerges.\\
 
The first model simply has a random force field $\vec{F}(\vec{r})$ which for simplicity we take Gaussian and time-independent with mean zero and covariance $$\langle F_i(\vec{r})F_j(\vec{r}')  \rangle = \delta_{i,j} f(\frac{\vec{r}-\vec{r}'}{\ell})$$ where $f$ is a rapidly decaying function  decay and $\ell$ is the correlation length. 
If we assume that the particle follows a straight trajectory (which is of course justified for short times), then the total momentum imparted by the force fields in time $\tau$, i.e. $\vec{P}_\tau= \int_0^\tau dt \vec{F}(\vec{r_0}+\vec{v}t)$ satisfies $\langle \vec{P}_\tau \rangle =0$ and the variance scales as 
\begin{equation}\label{eq: scaling variance}
\langle |\vec{P}_\tau|^2\rangle \propto  \frac{\tau v}{\ell} \times (\frac{\ell}{v})^2  \sim   \tau  \frac{\ell}{v} 
\end{equation}
by a simple central limit argument, where $ \frac{\tau v}{\ell}$ is roughly the number of independent values of the force field, seen in time $\tau$.  Comparing with \eqref{md} immediately gives $B(v)\sim 1/v$. That calculation promises to hold more generally when the force is continuous and random in an unbiased, translation-invariant way with short-ranged correlations as long as the central limit applies to the integral (written along the $i-$direction in which the particle moves at great speed $v$)
\begin{equation}
mv_i(\epsilon) - mv_i(0) = \int_0^\epsilon \id s\, F_i(vs) = \frac 1{v} \int_0^{v\epsilon} \id y \,F_i(y) = \sqrt{\frac{\epsilon}{v}} \,Z
\end{equation}
where the last equality is in distribution for $Z$ a Gaussian random variable with mean zero and finite variance. ($F_i(x)$ is the force at position $x$ with the particle starting at time $s=0$ from $x=0$ and moving with speed $v\gg 1$ for a small time $\epsilon$). Breaking of that central limit argument can occur when the force is conservative (gradient of a potential) and the randomness is on the level of the potential and not on the level of the force.  We have no specific model in mind for space plasmas.  A simple picture would take a dilute neutral plasma for charged particles, ions and electrons.  We take such a test particle and it seems reasonable to suppose then that the first two terms in \eqref{fknon} describe the evolution of its velocity distribution before a nonequilibrium source is added.  For that source (to be responsible for the last term in \eqref{fknon}) we assume an electromagnetic (EM) field which is stationary far-from-equilibrium. Here many details remain lacking, but it does not seem unreasonable to effectively think about a time-dependent Lorentz force acting on the test particle. Then, the electric field is not gradient by Faraday's law. Secondly we need that the EM-field is sufficiently mixing in the sense to allow sufficient spacetime decorrelation for the central limit argument above to apply.  That requires of course also that the variance of the force remains bounded in that nonequilibrium EM-distribution. Various discussions in the literature are of course devoted to that point.  It does remain a major problem in space plasma-physics to determine the nature of turbulent plasma that transfers energy from EM-fluctuations to the heating of plasma particles; see e.g. \cite{hol}.\\

The above complications invite a second more simple model, in the tradition of toy models in statistical mechanics as often demonstrated in the work of Christian Van den Broeck. Here we refer to his paper \cite{tay} where he discusses the phenomenon of Taylor dispersion \cite{tdif} in terms of a telegraph process \cite{kel,kac}. It gives a clear analogy with active (velocity) diffusion that leads to $B(v) \propto 1/v$.\\
In one dimension, if the amplitude of the force $F$ is constant but the sign flips between two values, we have an active diffusion in velocity space that mimics the run-and-tumble dynamics for active particles in one dimension, see e.g. \cite{act}, 
\[
\dot{v} = \eta\,F
\]
where $\eta = \pm 1$, flipping at rate $v/\ell$, where $\ell$ is a length scale related to the flipping of the force, be it a mean free path between collisions, a wave length or a distance like between magnetic mirrors.  We think here of a particle for which the sign of the force changes randomly at a rate that is proportional to its speed $v$.  Larger speeds brings the particle faster in a region (over distance $\ell$) where the force points in the opposite direction.  We get then a diffusion parameter $B(v) \simeq \ell\,F^2/v$, as enters the telegraph equation for the density function on speeds, \cite{tay,act}.\\
We believe that the two-dimensional model of active Brownian particles discussed in \cite{abi} provides even a further analogy (with the same conclusion) when the spatial coordinate there is replaced with the velocity, the harmonic trap is taken as Newtonian friction and a Lorentz force is added where the electric field undergoes rotational diffusion in the plane.\\
At any rate the behavior of $\gamma(v) \sim v^{-3}$ (from Coulomb scattering) and $B(v)\sim v^{-1}$ (from Taylor dispersion) are exactly what is needed to produce power-law decay ($\eta=1$) in the tails of the speed distribution.

\section{Injecting energy via negative friction}\label{gran}

As mentioned at the end of Section \ref{gen} we can also change the friction (instead of the diffusion in Section \ref{dac}).  Mathematically, the scenario is very similar but the physics for producing nonMaxwellians with heavy tails is now related to negative friction.  As mentioned in \cite{depo} Lord Rayleigh was among the very first in his ``The theory of sound'' (1877) to study motion with energy supply via friction.  energy producing high energy cosmic radiation. In the present section we discuss a specific example which is inspired by the physics of granular gases under vibration.  The vibration is here taken as coming from the oscillatory motion of a macroscopic object (the boundary) giving energy to the grain particles.\\

Driven granular gases have been widely studied theoretically and experimentally. The kinetics poses a wide range of interesting questions; see e.g. \cite{aba,kinb,and,san}.  One of the main phenomena is the deviation from Maxwellian speed distributions.  The literature is vast, and we only mention some references.
  For studies on establishing stretched exponentials, see e.g. \cite{bob} and \cite{ben}.   
 For the influence of correlations on the velocity statistics
of scalar granular gases, see \cite{abal,and}.
Many other people have generalized the result to $d>1$; see  \cite{mher}. A more recent review with many references and a treatment of inelastic Maxwell molecules is found in \cite{pra}. As far as we know there is no experimental evidence for power-law decay in the speed distribution.\\

One specific and simple dynamical mechanism to create fat tails in the velocity distribution of granular gases goes as follows. Consider ``pseudo-Maxwell molecules'' where the collision rate between the grains is governed by a restitution coefficient that depends on the relative speeds; see also \cite{pra} and the references there for similar and more detailed considerations.  In other words, the elasticity of the energy exchanges depends on the speed.  As a (one-dimensional) model we think of
particles  where in collision the velocity is rather drastically reset to close to zero during those (inelastic) collisions (events of resetting the speed to zero), happening after a random time as ticked by an exponential clock at rate $\lambda(v)$. That parameter is the escape rate in a Markov jump process, and is physically (in the diffusion approximation) related to the friction parameter $\gamma(v) = m\lambda(v)$ of \eqref{fknon}; the bigger $\lambda(v)$ the easier energy is dissipated away.  On the other hand, between the resettings,  the particles are driven (e.g. via vibrations) to increase their speed so that in all we have a single particle driven stochastic dynamics for the speed $v_t \geq 0$ of the form \eqref{md},
\begin{equation}\label{eas}
\dot{v}_t  =  \frac{B(v_t)}{m^2v_t}\qquad \text{ and with resettting }\;\;
v_t  \rightarrow  0 \;\;\text{at rate  }\;\; \gamma(v_t)/m > 0
\end{equation}
To get an equilibrium Maxwellian density we would need to add thermal diffusion, but we ignore that here for simplicity.  Obviously now, if $B\equiv 0$, then the grains cool down and all speeds go to zero eventually.  Instead, we take an energy injection of the scaling form 
\begin{equation}\label{asg}
B(v) = k_BT\,\gamma(v) \,\left(\frac{v}{v_0}\right)^{2\eta},\qquad v/v_0\gg 1
\end{equation}
as in \eqref{eta}, where the temperature $T$ and the speed $v_0$ are reference values, and
\begin{equation}\label{asgg}
\gamma(v) = m\nu_0\,\exp\left(-\frac{c}{2\xi}\,\left(\frac{v}{v_0}\right)^{2\xi}\right),\qquad v\gg 1
\end{equation}
for parameters  $\xi>0,c\geq 0$, making a stronger high energy localization for $c>0$, and with reference frequency $\nu_0>0$. 
(Below we require that $\xi\leq 1-\eta$ and $\xi\downarrow 0$ is allowed. If $\xi+\eta=1$ then $c< mv_0^2/(k_BT)$ must hold.  For inelastic hard spheres, $c=0$ and constant $\gamma(v) = m\nu_0$ are certainly a good option.)\\
On the other hand it is easy to see that a stationary density $\rho_s(v)$ for \eqref{eas} satisfies
\begin{eqnarray}
\frac{\id}{\id v}\left(\frac{B(v)}{m^2\,v}\,\rho_s(v)\right) + \frac{\gamma(v)}{m}\,\rho_s(v) &=&0\nonumber\\
\frac{\id}{\id v}\log \rho_s &=& -\frac{m\gamma(v)\,v}{B(v)} - \frac{\id}{\id v}\log\left(\frac{B(v)}{m^2\,v\,v_0\nu_0}\right)\nonumber\\
\frac{\id}{\id v}\log \rho_s &=& -\frac{m\,v_0^{2\eta}}{k_BT}\,v^{1-2\eta} +(1 - 2\eta)\,v^{-1} + \frac{c}{v_0^{2\xi}}\,v^{2\xi-1}\label{ahk}
\end{eqnarray}
which we want to solve for large $v$.  Stretched exponential decay of the stationary speed density follows exactly like in \eqref{resu} whenever $\xi + \eta < 1$ in which case
\[
\rho_s \propto \exp\left[ -\frac{m\,v_0^{2}}{2k_BT(1-\eta)}\,\left(\frac{v}{v_0}\right)^{2(1-\eta)}\right]
\]
independent of $c\geq 0$.
The main ingredient therefore is again to estimate the coefficient $\eta$ in \eqref{asg} for the energy injection \eqref{md}.  There is however no unique answer here, depending on the experimental conditions (vibration frequency, particle shape, material).  We refer to e.g. \cite{aba,san,sn,eve} for various studies and similar conclusions.\\
Note that exactly such a model as above was used by Fermi in \cite{Fer}.  Indeed again $\kappa-$distributions appear in the above from \eqref{ahk} for the choice $B(v) = bv^2$ (with some $b>0$) in \eqref{eas} (which is Eq. 3 in \cite{Fer}) and $\eta=1, c=0$ in \eqref{asg}--\eqref{asgg} (where the resetting to a small speed $v_0$ is discussed in Section IV of \cite{Fer}).  Indeed sections \ref{dac}--\ref{plas} and \ref{gran} are two sides of the same coin. Here the physics of space plasmas and granular gases meet.  That is in particular true in the Fermi-Ulam model, \cite{ula}. 
 Such models have been extended e.g. to two dimensions for chaotic billiards with time-dependent boundaries, \cite{consi}.  In the light of the previous sections, the scenario of combining localization with energy injection, be it from stochastic acceleration or from negative friction, will generically lead to nonMaxwellian velocity distributions with suprathermal tails.

\section{Conclusions and outlook}

The observation of suprathermal tails in velocity distributions indicate that high energy levels get an unusual large weight compared to the equilibrium case, i.e., a case of population inversion in velocity space.  We combined two classical effects, localization and diffusive acceleration, to set up a scenario for space plasmas and their power law tails in the speed distribution. The diffusive acceleration is related to Taylor dispersion for which we discussed a relation with the telegraph process as introduced by Christian Van den Broeck.  For granular gases we attempted a model combining localization with energy injection (negative friction).  It would be interesting to add other examples, such as the dynamical generation of nonPlanckian laws \cite{arcade} or, even outside physics, for the understanding of the Pareto distribution of wealth in a society; see e.g. \cite{arn}.   For the last example and depending on the economy it may well be that the wealth runs away towards the rich via a phenomenon similar to diffusive acceleration, and cannot quite redistribute itself due to localization effects.\\
The subject of nonMaxwellian velocity distributions touches on a number of fundamental issues in physics, which more recently have been named localization and active diffusion and that relate to older subjects introduced by Rayleigh (in the theory of sound), Kelvin and Kac (for the telegraph process), Fermi (for the origin of cosmic radiation), Taylor (for the subject of turbulent diffusion) and Ulam (on statistical properties of dynamical systems).  That those fundamental topics of statistical mechanics can be discussed via technically simpler and unifying considerations and models has been a precious motif in many examples set by Christian Van den Broeck.

\end{document}